\documentclass[reprint,superscriptaddress,amsmath,amssymb,aps,pra]{revtex4-2}

\usepackage{amsfonts}
\usepackage{amsmath}
\usepackage{amsthm}
\usepackage{mathrsfs}
\usepackage{amscd}
\usepackage{amssymb}
\usepackage{subfigure}
\usepackage{amsxtra}
\usepackage{dcolumn}
\usepackage{bm}           
\usepackage{bbm}
\usepackage{graphicx}
\usepackage{epstopdf}
\usepackage{color}
\usepackage[dvipsnames]{xcolor}
\usepackage[colorlinks,linkcolor=Blue,citecolor=Green]{hyperref}

\newcommand{\la}{\langle}
\newcommand{\ra}{\rangle}

\newcommand{\cF}{\mathcal{F}}
\newcommand{\cZ}{\mathcal{Z}}

\newcommand{\Op}[1]{\hat{#1}}

\newcommand{\osigma}{\Op{\sigma}}

\newcommand{\oH}{\Op{H}}

\newcommand{\oZ}{\Op{Z}}

\newcommand{\tr}{\ensuremath{{\rm tr}}}

\newcommand{\uul}[1]{\ensuremath{\smash[b]{\underline{\underline{#1}}}}}

\usepackage{xcolor}

\begin{document}

\preprint{APS/123-QED}

\title{Criticality-Enhanced Precision in Phase Thermometry}
\author{Mei Yu}
\affiliation{Naturwissenschaftlich-Technische Fakult{\"a}t, Universit{\"a}t Siegen, Siegen 57068, Germany}

\author{H. Chau Nguyen}
\affiliation{Naturwissenschaftlich-Technische Fakult{\"a}t, Universit{\"a}t Siegen, Siegen 57068, Germany}

\author{Stefan Nimmrichter}
\affiliation{Naturwissenschaftlich-Technische Fakult{\"a}t, Universit{\"a}t Siegen, Siegen 57068, Germany}

\begin{abstract}
Temperature estimation of interacting quantum many-body systems is both a challenging task and topic of interest in quantum metrology, given that critical behavior at phase transitions can boost the metrological sensitivity.
Here we study non-invasive quantum thermometry of a finite, two-dimensional Ising spin lattice based on measuring the non-Markovian dephasing dynamics of a spin probe coupled to the lattice. We demonstrate a strong critical enhancement of the achievable precision in terms of the quantum Fisher information, which depends on the coupling range and the interrogation time. Our numerical simulations are compared to
instructive analytic results for the critical scaling of the sensitivity in the Curie-Weiss model of a fully connected lattice and to the mean-field description in the thermodynamic limit, both of which fail to describe the critical spin fluctuations on the lattice the spin probe is sensitive to. Phase metrology could thus help to investigate the critical behaviour of finite many-body systems beyond the validity of mean-field models. 
\end{abstract}
\maketitle

\section{Introduction}
Temperature is one of the key thermodynamical quantities that determine the feasibility and performance of quantum experiments and technological applications such as quantum sensors \cite{Degen2017} and quantum computers \cite{Nielsen2002,Ladd2010}. Its precise measurement is therefore crucial, but also extremely challenging in the often desired regime of low temperatures \cite{Wu1998,Mehboudi2015,Pasquale2016,Correa2017,Hovhannisyan2018}. 
To estimate the temperature of a quantum system, one brings in a probe, i.e., the quantum thermometer, couples it to the system and measures it. Repeating this procedure, either in parallel with independently prepared copies, or sequentially with the same continuously coupled probe, one infers the temperature from the statistics of measurement outcomes.  One can measure the probe after it has already achieved thermal equilibrium with the system \cite{Mehboudi2019,Stace2010,Correa2015,Pasquale2016,Correa2017,Mok2021,Mihailescu2023,Abiuso2023}, or one can monitor its non-equilibrium dynamics \cite{Hangleiter2015,Jevtic2015,Mancino2017,Hofer2017,Cavina2018,Bouton2020,Mitchison2020,Oghittu2022,Mitchison2022,Yuan2023, Boeyens2023,Zhang2022,Dajian2022}.

For the assessment and optimization of quantum thermometry protocols, two main paradigms are pursued in the literature. Either one characterizes and and maximizes the \textit{globally} achievable accuracy averaged over a range of possible temperature values, typically based on Bayesian estimation and data analysis methods \cite{Rubio2021, Mok2021, Boeyens2021, Glatthard2022, Rubi2023}. Or one focuses on the \textit{locally} achievable precision of the estimate around fixed temperatures, as described by the Cram\'er-Rao bound in the limit of asymptotic data \cite{Fraser1964, Goold2018, Mehboudi2019}. The main goal is then to optimize the measurement strategy with respect to the probe's quantum Fisher information (QFI) on temperature. This local paradigm is the most widespread one \cite{Stace2010, Correa2015, Mehboudi2015, Correa2017, Potts2019, Mitchison2020}, and we will adopt it here to investigate temperature estimation near the critical point of an interacting spin system.

Physical systems close to a quantum or classical phase transition are extremely sensitive to small fluctuations of the relevant parameters, which can be exploited for quantum metrology and enhanced sensing. Previous theoretical studies in critical quantum metrology were focused on quantum critical systems in the ground state as probes for the estimation of Hamiltonian parameters \cite{Rams2018, Garbe2020, Chu2021, Raffaele2023,Mihailescu2023b}, e.g., the magnitude of an external magnetic field, or intra-system coupling strengths. However, exploiting the phase transition for enhanced temperature estimation is still a open problem.

In principle, one can infer the temperature of an equilibrium system from energy measurements, but in a strongly interacting many-body system, this would require global multi-particle measurements that are difficult to implement without disturbing the system. In certain types of systems away from criticality, measurements of extensive observables could be a viable alternative \cite{Mehboudi2015}. Another one is the minimally invasive “local” probing of small subsystems in order to infer the global temperature \cite{Pasquale2016,Giacomo2017,Correa2022}. In any case, an exponentially bad sensitivity at low temperatures is expected both for local and global strategies, unless the probed system is gapless \cite{Hovhannisyan2018, Aybar2022}.

In this work, we study phase thermometry: a non-invasive temperature estimation strategy in which a probe is brought into contact with a localized part of the sample system and undergoes pure dephasing, such that the sample temperature can be estimated by measuring the probe coherence over time. 
Previous works have considered phase thermometry of large bosonic samples with Ohmic spectral densities \cite{Razavian2019, Yuan2023} as well as non-interacting Fermi gases \cite{Mitchison2020,Oghittu2022, Mitchison2022,Sindre2023}.

Here, we set out to explore temperature estimation of interacting systems close to a phase transition and investigate  phase thermometry of a two-dimensional Ising model at zero field \cite{friedli2017}. Ising spin lattices may represent various physical scenarios such as arrays of Rydberg atoms \cite{Gross2017,Bernien2017,Browaeys2016,Barredo2018}, systems of nitrogen vacancy (NV) centers \cite{Jelezko12004, Jelezko22004, Robledo2011, Aboebeih2018, Degen2021, Gulka2021}, a spin-polarized Fermi gas \cite{Cetina2015, Cetina2016}, and macroscopic magnetic systems \cite{friedli2017}.

A single additional spin immersed in the sample and interacting with a controlled range of surrounding sample spins will experience temperature-dependent dephasing and thus serve as the phase probe. We will show that its sensitivity is critically enhanced at the phase transition point, allowing us to read out a significant portion of the temperature information contained in the state of those surrounding sample spins. We compare our Monte-Carlo simulation results to two well-known effective descriptions: the Curie-Weiss model \cite{friedli2017} in which each spin couples to the uniform sample average of all other spins, and the mean-field approximation.
Both capture the critical sensitivity boost qualitatively, but they fail to predict both the critical point and the exact enhancement quantitatively. 

We structure the manuscript as follows: section \ref{local_estimation} briefly reviews how to characterise the temperature estimation precision by means of the Fisher information. In section \ref{probe_dyn}, we introduce our model for the phase thermometer immersed in a thermal Ising spin lattice, the time evolution of which we solve using a previously developed framework \cite{Yu2023}. We then present our main result, the quantitative assessment of thermometric performance in terms of the Fisher information around the critical point, in section \ref{thermometry_spin_sys}. It starts with the effective Curie-Weiss model, which is analytically tractable in the thermodynamic limit, followed by the exact numerical treatment of a finite zero-field Ising lattice of $20\times 20$ spins and the associated mean-field approximation. The scaling of the Fisher information is evaluated as a function of the number of lattice spins the probe interacts with, and we also consider a high-temperature expansion to approximate the behaviour in the paramagnetic regime far above the critical temperature. Finally, we conclude in section \ref{conclusion}. 

\section{Precision of temperature estimates} \label{local_estimation}

Quantum estimation protocols generically start from a quantum system initialized in a fiducial state $\rho_p(0)$, which picks up information about a parameter, here the temperature $T$ of a thermal environment, as it evolves to a state $\rho_p(t,T)$ at time $t$. One then performs a measurement, described by a positive operator-valued measure (POVM) $\bigl\{ \coprod_{x}\bigl\}$ satisfying $\int dx \coprod_{x} = \openone$, which results in a random outcome $x$ at the likelihood $p(x|T)= \tr \{ \rho_p(t,T) \coprod_x\}$. 
The protocol is repeated $M$ times to gather an appropriately large data sample $\mathbf{x} = \{x_1, x_2, ..., x_M\}$, from which one can construct a temperature estimator $\tilde{T}$ yielding the estimate $\tilde{T}(\mathbf{x})$. One speaks of an unbiased estimator when the expectation value of the estimator over the likelihood of outcomes at a given true value $T$ matches the true value, $\mathbb{E}[\tilde{T} (\mathbf{x})] = T$ \cite{Fraser1964, Rubio2021}. 
This may hold merely approximately or asymptotically, in the limit of large data, as often the case with the commonly used maximum-likelihood estimator, for example.

In equilibrium thermometry, the system is allowed to exchange energy with the environment until it equilibrates to a thermal state after a sufficiently long time, $t\to \infty$. One then measures in the energy basis to infer its thermal populations and thereby estimate the temperature. In phase thermometry, a probe system is prepared in a superposition state that is then subjected to a thermal sample inducing dephasing over time, which one detects by performing phase-sensitive interferometric measurements on the probe at finite times $t$. We will focus on this case here, but for comparison, we will also consider directly measuring the equilibrium state of a number of spins in the thermal sample.

The statistical uncertainty of the temperature estimate can be defined with help of a suitable \emph{relative} error quantifier. This allows us to compare the achievable precision at different temperature scales. A suitable common choice is the relative mean-square deviation from the true value, $\mathbb{E} [(\tilde{T}-T)^2]/T^2$, which for unbiased estimators can be obtained from the data as the variance divided by the squared mean, $\mathrm{Var}[\tilde{T}]/(\mathbb{E}[\tilde{T}])^2$. This figure of merit for the local precision of temperature estimates at a given true temperature $T$ obeys the (quantum) Cram\'{e}r-Rao bound \cite{Braunstein1994,Cramer1999},
The statistical uncertainty of the temperature estimate can be defined with help of a suitable \emph{relative} error quantifier. This allows us to compare the achievable precision at different temperature scales. A suitable common choice is the relative mean-square deviation from the true value, $\mathbb{E} [(\tilde{T}-T)^2]/T^2$, which for unbiased estimators can be obtained from the data as the variance divided by the squared mean, $\mathrm{Var}[\tilde{T}]/(\mathbb{E}[\tilde{T}])^2$. This figure of merit for the local precision of temperature estimates at a given true temperature $T$ obeys the (quantum) Cram\'{e}r-Rao bound \cite{Braunstein1994,Cramer1999},
\begin{equation}\label{eq:CRB}
    \frac{\mathbb{E}\big[\tilde{T}(\mathbf{x})-T)^2\big]}{T^2} \geq \frac{1}{M T^2 \cF_T} \geq \frac{1}{M T^2 \cF^Q_T}.
\end{equation}
Here, the first inequality is specific to the chosen quantum measurement and given in terms of the Fisher information (FI) of the associated likelihood with respect to temperature,
\begin{equation}
    \cF_T= \int dx \, p(x|T)\left[ \frac{\partial \ln p(x|T)}{\partial T} \right]^2. \label{FI_equ}
\end{equation}
It quantifies the sensitivity of the likelihood to small temperature changes around $T$.

The more fundamental second inequality is obtained by optimizing the FI over all possible POVMs, $\cF^Q_T = \max_{\{ \coprod_x\}} \cF_T$. The so defined quantum Fisher information (QFI) is a function of the state $\rho_p(t,T)$ and sets the ultimate precision bound, i.e., the maximum local temperature sensitivity, attainable for any chosen measurement on that state. 

The QFI can be given explicitly in terms of the symmetric logarithmic derivative (SLD) operator $\hat{L}_T$, which is the self-adjoint solution to the Lyapunov equation
\begin{equation}
   \frac{\partial\rho_p(t,T)}{\partial T} = \frac{1}{2} \left[\rho_p(t,T) \hat{L}_{T} + \hat{L}_{T} \rho_p(t,T)\right]. \label{SLD_eq}
\end{equation}
The solution depends on the probe state, and its eigenbasis gives the corresponding optimal projective measurement   \cite{Paris2009}. The QFI then reads as
\begin{equation}
    \cF^{Q}_T = \tr \left[ \rho_p(t,T) \hat{L}_{T}^2\right] = 2 \sum_{nm} \frac{|\langle \psi_m|\partial_T \rho_p(t,T)|\psi_n \rangle|^2}{\rho_n + \rho_m}. \label{qfi_eq}
\end{equation}
Here, the second expression follows by expanding the probe state in its eigenbasis, $\rho_p(t,T) = \sum_n \rho_n | \psi_n\rangle \langle \psi_n |$, which leads to
\begin{equation}
    \hat{L}_T = 2 \sum_{nm}\frac{\langle \psi_m|\partial_T \rho_p(t,T)|\psi_n \rangle}{\rho_n + \rho_m}|\psi_m \rangle \langle \psi_n |.
\end{equation}

In the following, we will conveniently express the temperature in terms of the inverse temperature scale parameter $\beta = 1/k_B T$. This reparametrization leaves the relative Cram\'{e}r-Rao bounds in \eqref{eq:CRB} invariant, $T^2 \cF_T = \beta^2 \cF_\beta$ and thus $T^2 \cF^Q_T = \beta^2 \cF^Q_\beta$, as one can easily check using \eqref{FI_equ}. These dimensionless bounds will be used to characterise the performance of our phase thermometer protocol.

\section{Zero-field Ising model with a single phase probe} \label{probe_dyn}

\begin{figure}[t]
    \centering
    \includegraphics[width=8.5cm]{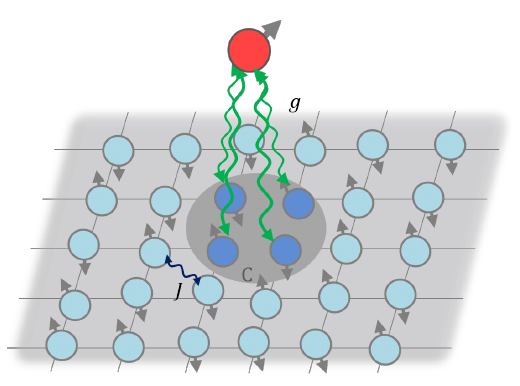}
    \caption{Two dimensional Ising lattice model. The probe spin is depicted on top of the lattice layer and interacts with system spins inside of a cluster $C$ (dark gray circle) with uniform coupling strength $g$. The system spins interact with each other at coupling strength $J$.} \label{sketch_graph}
\end{figure}

We consider a finite two-dimensional Ising spin lattice with nearest-neighbour interactions as the thermal sample whose temperature we seek to estimate. It is described by the Hamiltonian
\begin{equation}
    \oH_s = - J \sum_{\langle ij\rangle} \hat{\sigma}_z^i \otimes \hat{\sigma}_z^j - h \sum_{i=1}^N \hat{\sigma}_z^i, \label{gene_Hamiltoian}
\end{equation}
where $N$ is the total spin number, $h$ the single-site energy due to an external field, $J$ the nearest-neighbour coupling energy, and $\la ij\ra$ denotes any possible pair of nearest neighbours assuming periodic boundary conditions. The dimensionless spin operator $\hat{\sigma}_z^i$ describes the Z-component of a spin-$1/2$ on the $i$-th lattice site with eigenvalues $m_i = \pm 1$. We will focus on the zero-field case $h=0$ and on $J>0$ in the following, which exhibits a transition between a ferromagnetic and a paramagnetic phase. We describe the sample in thermal equilibrium at temperature $T$ by the Gibbs state $\rho_s = e^{-\beta \oH_s}/\cZ$, with the partition function $\cZ = \tr ( e^{-\beta \oH_s} )$.

Even though the above Ising spin lattice Hamiltonian is a highly simplified model for ferromagnetism, it can be representative of a large class of systems (including realistic ones) and admit a quantitative understanding of these systems near their critical point \cite{friedli2017}. This is because, by virtue of universality, the behavior of a system in the vicinity of a phase transition becomes essentially independent of its microscopic details.

As our non-invasive phase probe, we consider a single spin-$1/2$ (or two-level system) that interacts with some lattice spins in its vicinity through the Ising Z-Z coupling; see Fig.~\ref{sketch_graph} for a sketch. For simplicity, we restrict the probe-sample interaction to a finite cluster of $n$ lattice spins within a certain radius and assume a uniform coupling rate $g$. This will allow us to compare the temperature information acquired by the probe with the information one could obtain if one were to directly measure the local configuration of the $n$ lattice spins. The joint probe-sample Hamiltonian is
\begin{eqnarray}
    \oH = \oH_s + \frac{\hbar \omega_p}{2}  \hat{\sigma}^{p}_z + \frac{\hbar g}{2} \hat{\sigma}^{p}_z \otimes \oZ_n, 
\end{eqnarray}
with $\hbar \omega_p$ the probe energy, $\hat{\sigma}_z^p$ the Pauli-Z matrix of the probe (with eigenvalues $\pm 1$), and $\oZ_n =\sum_{i=1}^n \hat{\sigma}^i_z$ the total spin of the $n$-site cluster. In the extreme case $n=N$, the probe senses the collective fluctuating magnetization of the whole sample.

The probe-sample coupling commutes with the probe Hamiltonian, which implies that the diagonal terms of the probe density operator in the energy basis remain constant. A generic initial probe state that is uncorrelated to the Gibbs state $\rho_s$ of the sample, 
\begin{equation}
    \rho_p = \begin{pmatrix}p & c \\ c^* & 1-p\end{pmatrix},
\end{equation}
will undergo pure dephasing due to averaging over the random phase shifts caused by the thermally occupied spin configurations of the sample. The reduced probe state evolves as
\begin{eqnarray}
    \rho_p(t) &=&  \tr_s \left[ e^{-i\oH t/\hbar}\rho_p \otimes\rho_s e^{i\oH t/\hbar}\right] \nonumber \\ 
    &=& \begin{pmatrix}p & c e^{-i\omega_p t}r(t,\beta)   \\ c^* e^{i\omega_p t} r^*(t,\beta) & 1-p\end{pmatrix}, 
    \label{probe_state}
\end{eqnarray}
with the decoherence factor
\begin{equation}
    r(t,\beta) = \tr [\rho_s e^{-i g \oZ_n t}] = \frac{1}{\mathcal{Z} } \tr [e^{-\beta \oH_s} e^{-i g \oZ_n t}].
    \label{charact_func}
\end{equation}
This factor has the form of a characteristic function that generates the moments of the $n$-site cluster's magnetization $\oZ_n$, via $\la \oZ_n^m \ra = (i/g)^m\partial_t^m r(0,\beta)$. Hence, one could in principle infer the full spin distribution of the cluster by sampling the decoherence factor at many different times. 

Here, we will be concerned with the maximum information about the sample temperature the probe can acquire. Starting from a maximally phase-coherent initial state, $\rho_p = |+\ra\la +|$ with $p,c=1/2$, the QFI of the probe state at an interrogation time $t$ with respect to $\beta$ becomes
\begin{eqnarray}
    \cF_\beta^Q (t) &=& \frac{4|\partial_{\beta} r|^2 + \big[r^* \partial_{\beta} r - r\partial_{\beta} r^* \big]^2}{4(1-|r|^2)}. \label{eq:QFI_probe}
\end{eqnarray}
For the results presented in the following, we evaluate the maximum of this QFI with respect to $t$, showing the highest possible phase thermometry precision in our setting. 

Crucially, our treatment is exact at this point; it encompasses effects such as coherence revivals, which go beyond Markovian dephasing models in the weak-coupling limit \cite{Breuer2002}. 
For an efficient numerical modelling, we make use of a binary array-based representation of spin networks \cite{Yu2023}: Let the binary array $\underline{b} \in \{0,1\}^{\times N}$ and the associated state vector $|\underline{b} \ra$ represent a specific $N$-spin configuration of the sample lattice, such that $\hat{\sigma}_z^i|\underline{b}\ra = (1 - 2 b_i) |\underline{b}\ra$ for all $i$. Short- or long-range spin-spin interactions can then be described in terms of adjacency matrices. In particular, let $\uul{\Gamma}$ be the $N\times N$ adjacency matrix of the lattice, taking the value $\Gamma_{ij}=-J$ for nearest-neighbour pairs $\la ij\ra$ and zero otherwise. The energy of a given spin configuration is then $E(\underline{b}) = (2h+8J)\sum_{i=1}^N b_i -2 \underline{b}\cdot \uul{\Gamma} \underline{b}$, and its thermal occupation probability is $p(\underline{b}) = e^{-\beta E(\underline{b})}/\cZ$.

Now, let us split the binary array, $\underline{b} = (\underline{b}',\underline{b}'')$, into the configuration $\underline{b}'$ of the $n$-spin cluster interacting with the probe and the configuration $\underline{b}''$ of the remaining $N-n$ spins. Given the $1\times n$ adjacency matrix $\underline{\Gamma}'$ of the probe-sample interaction, which in our uniform coupling model amounts to simply $\underline{\Gamma}' = g (1,1,\ldots,1)$, the decoherence factor reads as
\begin{equation}
    \begin{split}
       & r(t,\beta) = \sum_{\underline{b}'} \left(\sum_{\underline{b}''}p(\underline{b}',\underline{b}'')\right) e^{-i t \underline{\Gamma}'\cdot \underline{b}'}. \label{charact_func_ad_matrix}
        \end{split}
\end{equation}

\section{Temperature sensitivity near the critical point} \label{thermometry_spin_sys}

The theoretical model at hand, we will now assess the maximum sensitivity of a phase probe to the temperature of an Ising lattice sample of $N$ spins at thermal equilibrium. Different approximate treatments will be compared to an exact simulation. Our aim is to scrutinize and quantify our expectation that the temperature sensitivity will reach a pronounced peak, and even diverge for $N\to\infty$, at the critical temperature of the ferromagnetic-paramagnetic phase transition.

We start with the extreme case of probing the whole sample magnetization ($n=N$) and introduce the instructive Curie-Weiss model, which serves as a form of mean-field approximation of the Ising lattice. The model can also represent a scenario of its own: a fully and uniformly connected spin network, see Fig.~\ref{curieweiss_sketch_graph}.

Subsequently, we will evaluate the exact results for the Ising lattice and varying cluster size $n$ and compare them to the sensitivity predicted by the conventional mean-field approximation. Moreover, we will consider a high-temperature expansion to model the thermometric precision in the paramagnetic regime.

\subsection{Curie–Weiss model}\label{sec:CW}

\begin{figure}[t]
    \centering
    \includegraphics[width=4.0cm]{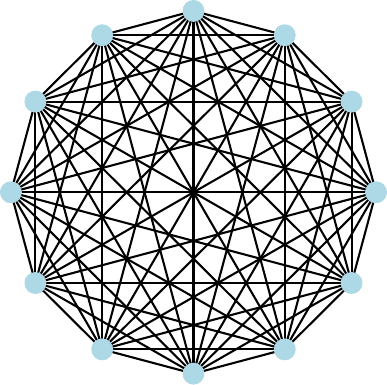}
    \caption{In the Curie–Weiss model, every spin interacts with all the others uniformly, as can be represented by a complete graph. In this graph, each spin is represented by a vertex. An edge connecting a pair of vertices $i$ and $j$ represents the coupling contribution $-(J/N)\hat{\sigma}_z^i \otimes \hat{\sigma}_z^j$ to the total energy, where $J$ represents the interaction strength and $N$ denotes the total number of spins (here: $12$).} \label{curieweiss_sketch_graph}
\end{figure}

The Curie-Weiss model is based on the idea that, in a ferromagnetic lattice, all spins align in response to a  collective magnetization or to an external magnetic field \cite{friedli2017}.
Rather than nearest-neighbour or more general site-specific couplings, each spin $\hat{\sigma}_z^i$ in the lattice shall react only to the collective spin of the whole lattice, $\sum_j \hat{\sigma}_z^j$. For consistency with the microscopic two-spin coupling model of strength $J>0$, we divide by $N$ and thus couple each spin to a lattice-averaged macroscopic spin density, described by the Hamiltonian  
\begin{equation}
     \oH_s^{\rm CW} = -\frac{J}{2N} \sum_{i,j =1}^{N} \hat{\sigma}_z^i \otimes \hat{\sigma}_z^j - h \sum_{i=1}^{N} \hat{\sigma}_z^i. \label{eq:CW_Ham}
\end{equation}
The factor $1/2$ prevents double counting of spin pairs, and the diagonal terms ($i=j$) in the double sum merely give an irrelevant energy offset. One can envision this model as a spin network defined by a complete graph with $N$ vertices, wherein every pair of distinct vertices is connected by an edge, as illustrated in Fig. \ref{curieweiss_sketch_graph}. We will restrict to the zero-field case ($h=0$) later, but all the following derivations also apply for $h \neq 0$. 

At thermal equilibrium, the partition function of the model is 
\begin{eqnarray}
     \mathcal{Z} &=& \sum_{m_1=-1}^{1}\ldots \sum_{m_N=-1}^{1} e^{ (J/2N) \beta \left( \sum_{i=1}^N m_i \right)^2 + h\beta \sum_{i=1}^N m_i } \nonumber \\ 
     &=& \sum_{M=-N}^{N} \binom{N}{\frac{N+M}{2}} e^{(J/2N) \beta M^2  + h\beta M } ,
     \label{part_funct_dis}
\end{eqnarray}
where $M = \sum_{i=1}^N m_i$ is the total dimensionless spin or magnetization. For a large lattice, $N\to \infty$, we can approximate the discrete sum over $M$ by an integral over a continuous magnetization density $m=M/N \in [-1,1]$ per number of spins,
\begin{equation}
    \mathcal{Z} \approx \int_{-1}^{1} e^{-N f(m)} dm \label{z_int},
\end{equation}
where the free energy density is approximately given as 
\begin{equation}
f(m) \approx 
\frac{1+m}{2} \ln \frac{1+m}{2} + \frac{1-m}{2} \ln \frac{1-m}{2} - \frac{\beta Jm^2}{2} -\beta h m ,   
\end{equation}
by virtue of the Stirling formula. 
The integral in Eq.~(\ref{z_int}) is dominated by the vicinity of the saddle point $m_0$ of the function $f$ where $f'(m_0)=0$ \cite{friedli2017}, 
which satisfies
\begin{equation}
    \tanh{\left( J\beta m_0 + h\beta \right)} = m_0 \label{saddle_point_eq}.
\end{equation}
We can thus perform the saddle point approximation, 
\begin{equation}
    \mathcal{Z} \approx e^{-N f(m_0)}\sqrt{\frac{ 2 \pi}{N f''(m_0)}},
\end{equation}
with $f''(m_0) = 1/(1-m_0^2) - J\beta$.

The saddle point solution $m_0$ of \eqref{saddle_point_eq} also constitutes the mean magnetization, $\la \sum_i \hat{\sigma}_z^i \ra = \beta^{-1}\partial_h \ln \cZ \approx N m_0$, and thus describes the spontaneous symmetry breaking at the ferromagnetic-paramagnetic phase transition. For $h=0$, spontaneous finite magnetization occurs at the critical point $J\beta_c = 1$ \cite{stanley1987,friedli2017}. 

\subparagraph{Phase thermometry}

For temperature sensing, we consider a qubit probe interacting with all $N$ lattice spins with uniform coupling rate $g$. The decoherence factor \eqref{charact_func} can then be given consistently in the continuum and saddle-point approximation for $N\to \infty$ as
\begin{eqnarray}
   r(t,\beta) 
   &=& \sum_{M=-N}^{N} \binom{N}{\frac{N+M}{2}} e^{(J/2N) \beta M^2  + (h\beta-igt) M } \nonumber \\
   &\approx& e^{-i\tilde{g}t m_0 - (\tilde{g}t)^2/ [2N f''(m_0)]}, 
   \label{dec_fact}
\end{eqnarray}
with $\tilde{g} = gN$. The effect of the lattice is a frequency shift proportional to the mean magnetization and, to second order in $\tilde{g}t$, a dephasing due to the thermal spin fluctuations with a characteristic ($e^{-1/2}$) coherence decay time scale $\tilde{g} \tau(\beta) = \sqrt{N f''(m_0)}$. The coherence decays more rapidly the greater the lattice, as $\tau(\beta) \propto 1/\sqrt{N}$.

\begin{figure}[t]
    \centering
    \includegraphics[width=8.5cm]{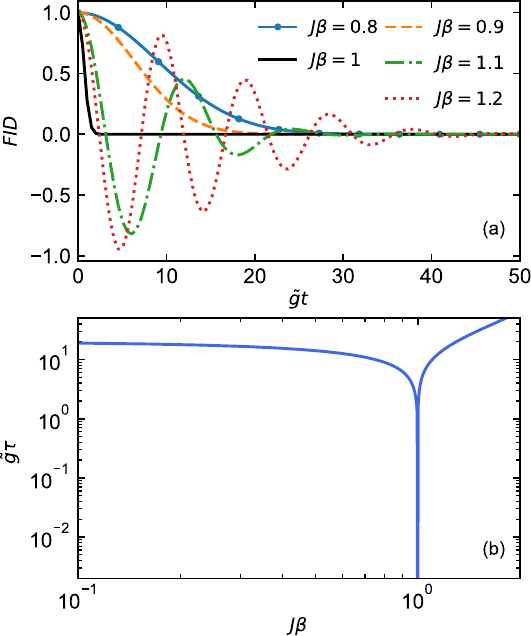}
    \caption{(a) Coherence dynamics of the spin probe interacting with the Curie-Weiss lattice at different temperatures over time, measured in terms of the free induction decay \eqref{eq:FID}. (b) Coherence decay time versus inverse temperature, as predicted by the Curie-Weiss model in the thermodynamic limit. We set the parameters $N = 400, h = 0$. \label{fig:cohe_secderi_plot}}
\end{figure}

To illustrate the behaviour, consider the probe initialized in the X-eigenstate $|+\ra$ (which is the optimal initial state for phase thermometry \cite{Mitchison2020}) and an observation of the spin-X component as a function of time. The expectation value of this coherence observable shows oscillations and the so-called free induction decay (FID) \cite{Mohamad2021},
\begin{equation} 
    \mathrm{FID}(t) = \tr \{\hat{\sigma}_x^p \rho_p(t)\} 
    = \cos (\tilde{g}t m_0) e^{-t^2/2\tau^2(\beta)}.\label{eq:FID}
\end{equation}
We plot this quantity at zero field and $N=400$ for various temperatures around the critical point in Fig.~\ref{fig:cohe_secderi_plot}(a). We observe a Gaussian decay for above-critical temperatures in the paramagnetic phase ($J\beta < 1$) where the mean magnetization vanishes ($m_0=0$).  Conversely, in the ferromagnetic phase, we observe a damped oscillation with a Gaussian envelope. The decay time is minimal at the critical point (black solid line) and increases further away from the critical point in both directions, see also Fig.~\ref{fig:cohe_secderi_plot}(b). In the limit of zero temperature, we have no decay, $m_0 \to 1$ and $\tau(\beta \to \infty) \to \infty$, as expected. In the paramagnetic phase, we always have $m_0 =0$, and in the high-temperature limit, $\tau(0) = 1/g\sqrt{N}$. In the ferromagnetic phase close to the critical point, we can Taylor-expand the $\tanh$ in \eqref{saddle_point_eq} in small $m_0>0$ and $\epsilon = 1-J\beta = 1-\beta/\beta_c<0$, resulting in $m_0 \approx \sqrt{-3\epsilon}$.

\subparagraph{Temperature precision at criticality}

The high sensitivity of the probe coherence to the thermal fluctuations of the sample around the critical point suggests that temperature could be probed more precisely. This is quantified by the QFI $\cF_\beta^{Q}$; the greater it is at a given temperature, the higher the estimation precision and the lower the uncertainty around that temperature. Inserting the decoherence factor \eqref{dec_fact} into the QFI \eqref{eq:QFI_probe} and several steps of algebra result in
\begin{eqnarray}
     \cF_\beta^Q &=& \frac{J^2}{4}\frac{[(1-m_0^2)^2f''(m_0)-2m_0^2]^2 }{(1-m_0^2)^6 [f''(m_0)]^4} \frac{t^4/\tau^4(\beta)}{e^{t^2/\tau^2(\beta)}-1}  \nonumber \\
     &&+ NJ^2\frac{m_0^2}{f''(m_0)}\frac{ t^2}{\tau^2(\beta)} e^{-t^2/\tau^2(\beta)},
     \label{QFI_Curie}
 \end{eqnarray}
In the paramagnetic phase ($\beta < \beta_c = 1/J$), where $m_0=0$ and $f''(0) = 1-\beta/\beta_c = \epsilon>0$, the QFI simplifies to
\begin{equation}
    \cF_{\beta < \beta_c}^Q = \frac{J^2}{4\epsilon^2} S \left(\frac{\tilde{g}^2 t^2}{N\epsilon} \right), \quad S (x) = \frac{x^2}{e^x-1}. \label{eq:QFI_CW_para}
\end{equation}
Optimal sensitivity is achieved at the time $t$ that maximizes the function $S$, which yields $\max_t \cF_{\beta\leq \beta_c}^Q = 0.162 J^2/\epsilon^2$ when $\tilde{g}^2t^2 = 1.594 N \epsilon $. 

In the ferromagnetic phase near the critical temperature, the magnetization is also small, $m_0 \approx \sqrt{-3\epsilon} \ll 1$, and we have $f''(m_0) \approx -2\epsilon$. Hence, to leading order in the temperature deviation $\epsilon<0$, 
\begin{equation}
    \cF_{\beta > \beta_c}^Q \approx \frac{J^2}{\epsilon^2} S \left(-\frac{\tilde{g}^2 t^2}{2N\epsilon} \right). \label{eq:QFI_CW_ferro}
\end{equation}
In total, this leads to the following critical exponents: the optimal QFI diverges like $\epsilon^{-2}$ and the optimal time before measuring the probe scales like $t \propto |\epsilon|^{1/2}$ for $\epsilon \to 0$ in the thermodynamic limit. The values remain finite for finite $N$. 

It is instructive to compare the temperature sensitivity of the phase spin probe to directly probing the local magnetization on any single lattice spin. This amounts to measuring $\hat{\sigma}_z^i$ at a random lattice site $i$. In the thermodynamic limit, the probability to obtain either outcome $\pm 1$ is given by $p_\pm = (1\pm m_0)/2$, and hence the FI of this measurement with respect to $\beta$ is
\begin{equation}
    \cF_{\beta}^{\rm (lc)} = \frac{(\partial_\beta m_0)^2}{1-m_0^2} \xrightarrow[\epsilon < 0]{\epsilon \to 0} -\frac{3J^2}{4\epsilon}.
\end{equation}
Above the critical point, both $m_0$ and the local FI vanish, whereas closely below the critical point, the FI diverges like $|\epsilon|^{-1}$ -- more weakly than the QFI of the phase probe.

\begin{figure}[t]
    \centering
    \includegraphics[width=8.5cm]{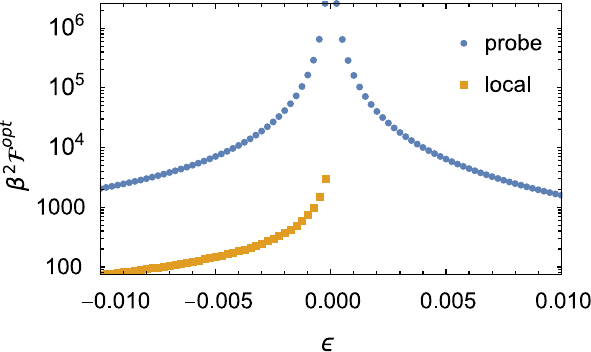}
    \caption{Optimal QFI of the phase probe (optimized over the interrogation time) for the Curie-Weiss model in the thermodynamic limit, plotted as a function of the relative temperature deviation $\epsilon$ from the critical point. 
     The right half ($\epsilon > 0$) represent the paramagnetic regime of $\beta < \beta_c$ and zero mean magnetization. The left half marks the ferromagnetic regime of low temperatures, where the direct local measurement of a single lattice spin yields non-vanishing temperature information. We assume a probe-lattice coupling strength $\hbar g/J = 0.4$ and the lattice coupling $J = 1/4$ in units of an arbitrary reference energy scale. \label{fig:Opt_QFI_Plot}}
\end{figure}

We illustrate the difference between the local FI and the optimal phase probe QFI close to the phase transition in Fig.~\ref{fig:Opt_QFI_Plot}. In both cases, only a single spin degree of freedom is measured, but the phase probe interacts with the whole lattice and thus picks up correlated thermal fluctuations around the mean magnetization $m_0$ through the dephasing factor \eqref{dec_fact}, leading to a much higher temperature sensitivity. In the following, we will see a similar, albeit less pronounced, advantage of the phase probe in the Ising lattice.

\subsection{2D Ising model at zero field}

We now turn to phase thermometry of the finite ferromagnetic Ising lattice at zero field, as described by the Hamiltonian \eqref{gene_Hamiltoian} with $h=0$ and $J>0$. To obtain our results and sample the thermally occupied spin configurations, we performed a Monte Carlo simulation on a $20 \times 20$ lattice ($N=400$) with periodic boundary conditions. The probability $p(\underline{b})$ of each spin configuration is simulated by the relative frequency over $1.2 \times 10^
7$ instances. The decoherence factor of the phase probe \eqref{charact_func_ad_matrix} can then be used to compute the QFI \eqref{eq:QFI_probe} at arbitrary times and its maximum $\cF_\beta^{\rm opt} = \max_t \cF_\beta^Q (t)$ with respect to time. 

\begin{figure*}
    \centering    
    \includegraphics[width=1.0\textwidth]{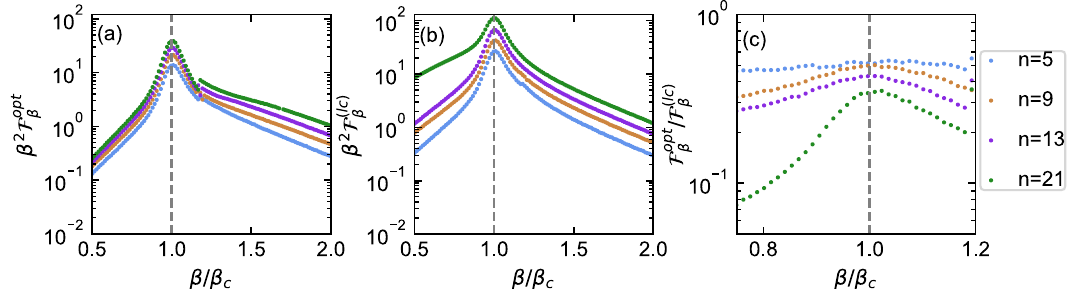}
    \caption{(a) Optimal QFI of the phase probe (optimized over time) interacting with spin clusters of varying size $n$ in a $20\times20$ Ising lattice, as a function of inverse temperature relative to the critical point (vertical dashed line). (b) FI associated to a direct local measurement of the configuration of the $n$ cluster spins. (c) Ratio of optimal probe QFI and direct local FI close to the critical point. We use the probe-cluster coupling strength $\hbar g/J = 0.4$ and set $J = 1/4$ (in arbitrary reference units). \label{local_pb_complt}}
\end{figure*}

Figure \ref{local_pb_complt}(a) shows the simulation results for the maximum QFI in relative units as a function of inverse temperature $\beta$, for various cluster sizes $n$ the probe interacts with. The inverse temperature is plotted relative to the known critical value $\beta_c = \ln{(1+\sqrt{2})}/2J$ for a 2D Ising model in the thermodynamic limit \cite{Onsager1994}, which matches the QFI peaks in this $20\times 20$ simulation. The visible jump at  $\beta \sim 1.2 \beta_c$ is a numerical instability detecting the spontaneous symmetry breaking in the ferromagnetic phase. At this point, the zero-field Ising system may randomly assume one of two opposite ferromagnetically ordered configurations that are degenerate in energy \cite{Aro2019}. 
The sudden change from a probabilistic mixture of both options to one of them directly affects the complex phase of the decoherence factor \eqref{dec_fact}, resulting in the observed jump. 

Although the phase thermometer probes the occupied configurations of an $n$-spin cluster, the measurement outcome is still a single bit value. Let us compare this to a direct measurement of the local cluster configurations $\underline{b}'$ in the Ising lattice, thermally occupied with the marginal probability $p_n(\underline{b}') = \sum_{\underline{b}''} p(\underline{b}',\underline{b}'') $. The temperature sensitivity of this $n$-bit measurement is given by the FI
\begin{eqnarray}
    \cF_{\beta}^{\rm (lc)} &=& \sum_{\underline{b}'} \frac{1}{p_n(\underline{b}')} \left[ \frac{\partial p_n(\underline{b}')}{\partial \beta} \right]^2 \nonumber \\
    &=& \sum_{\underline{b}'} \frac{1}{p_n(\underline{b}')} \left[\sum_{\underline{b}''} E(\underline{b})p(\underline{b})\right]^2 - \la \oH_s\ra^2, \label{FI_localeq}
\end{eqnarray}
which we could sample from our Monte-Carlo simulation for moderate $n \lesssim 30$. 
The results are plotted in Fig. \ref{local_pb_complt}(b) for the same temperature range and cluster sizes as before. We once again observe a peak at the phase transition, but now a less steep decay into the paramagnetic phase. Naturally, the FI of this $n$-bit measurement is greater than the QFI of the phase probe, but the ratio of both quantities close to the critical point grows only sublinearly with $n$, as can be seen in (c). Deep in the ferromagnetic phase, in particular, the lattice spins are strongly aligned (i.e., correlated) such that the phase probe picks up almost all the temperature information in the cluster. Conversely, in the paramagnetic phase, the correlations are broken and the disparity between both measurements grows.

\begin{figure}[htp]
    \centering   
    \includegraphics[width=8.5cm]{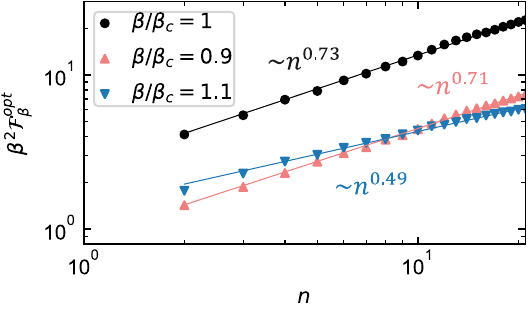}
    \caption{Scaling of the optimal probe QFI with the size $n$ of the lattice spin cluster that it interacts with, using the parameters of Fig.~\ref{local_pb_complt}. We compare the scaling at the critical point to the scaling at a $10\%$ higher temperature ($\beta = 0.9\beta_c$) and at a $10\%$ lower temperature. The solid lines are linear fits on the double-logarithmic scale with slopes $0.73$, $0.71$, and $0.49$, respectively. 
    \label{opt_qfi_n}}
\end{figure}

The scaling of the optimal QFI with cluster size $n$ is plotted on a double-logarithmic scale in Fig.~\ref{opt_qfi_n} at the critical point, as well as $10\,\%$ above and below. From the average slopes, we obtain a critical scaling with $n^{0.73}$; weaker in the ferromagnetic phase and about as strong in the paramagnetic phase. This sublinear scaling illustrates the strong correlations among the cluster spins. 

In the following, we investigate approximate analytical models for the temperature sensitivity based on the widely used mean-field theory (MFT) and on a high-temperature expansion (HTE) of the Gibbs state. We will see that these models yield accurate predictions for temperatures deep in the ferromagnetic and paramagnetic phase, respectively, but they fail to reproduce the critical behaviour of our exact numerical simulation.

\subparagraph{Mean-field approximation} 

In MFT, one describes a large ferromagnetic Ising lattice by a mean homogeneous magnetization $m_0 = \la \hat{\sigma}_z^i\ra$ $\forall i$ and expands in the microscopic fluctuations around this mean \cite{kadanoff2009}. Inserting the ansatz $\hat{\sigma}_z^i = m_0 + \delta \hat{\sigma}_z^i$ into the Hamiltonian \eqref{gene_Hamiltoian} at $h=0$ and neglecting second-order terms in the fluctuation operators, one obtains
\begin{eqnarray}
    \oH^{(\mathrm{MFT})}_s &\approx& -J\sum_{\langle i,j\rangle} m_0\left( \hat{\sigma}_z^i + \hat{\sigma}_z^j -m_0 \right)  \nonumber \\
    & =& \frac{N J q m_0^2}{2} - Jqm_0\sum_{i=1}^{N} \hat{\sigma}_z^i, 
    \label{Hamiltonian_mft}
\end{eqnarray}
with $q$ the number of nearest neighbours per spin, here $q=4$. The Hamiltonian describes independent spins aligning with respect to an effective mean field $h_{\rm MFT} = Jqm_0$, which itself is determined by the mean spin alignment. At thermal equilibrium, the partition function and the thermal occupation probability are products of $N$ identical single-spin terms,
\begin{eqnarray}
    \mathcal{Z} &=& e^{-\beta N q J m_0^2/2}\left[ 2\cosh{\left(Jqm_0 \beta \right)} \right]^N, \label{eq:Z_MFT} \\
    p(\underline{b}) &=& \frac{e^{-2 Jqm_0 \beta \sum_i b_i}}{(1+e^{-2 Jqm_0 \beta })^N}. \label{eq:pb_MFT}
\end{eqnarray}
The corresponding average alignment of each spin leads to the self-consistency equation $m_0 = \la \hat{\sigma}_z^i\ra = \tanh[Jqm_0\beta]$, which overestimates the critical temperature here to $J\beta = 1/q = 0.25$ (as opposed to $J\beta_c \approx 0.44$ for $N\gg 1$). The self-consistency condition has the same form as in the Curie-Weiss model in Sec.~\ref{sec:CW} if we replace $Jq$ by an average effective coupling strength, but the key difference is that we have a collective spin-$N/2$ model there and $N$ individual spins here, reflected by the distinct partition functions. 

The effect on the phase probe over time follows by inserting \eqref{eq:pb_MFT} into the decoherence factor expression \eqref{charact_func_ad_matrix}, 
\begin{equation}
    r(t,\beta) = \left[1 - \frac{1- e^{-igt}}{1+e^{2Jqm_0 \beta}} \right]^n \approx e^{\left(1-\cos gt -i\sin gt\right) np_1},
\end{equation}
with the single-spin excitation probability $p_1 = 1/(1+e^{2Jqm_0\beta})$. The approximation holds for moderate cluster sizes $n$ in the ferromagnetic phase. From this, we arrive at an analytic expression for the QFI,
\begin{eqnarray}
    \cF_{\beta}^{(\mathrm{MFT})} &=& \frac{J^2q^2n^2\left(m_0+  \beta \partial_{\beta} m_0 \right) ^2}{4 \cosh^4 (Jqm_0\beta)} e^{-2\gamma(1-p_1)} \nonumber \\
    &&\times\left[ \sin^2 (gt) + \frac{(1-\cos gt)^2)}{1-e^{-2\gamma(1-p_1)}}\right].\label{QFI_mft}
\end{eqnarray}
As the derivative $\partial_\beta m_0$ diverges at the (overestimated) critical temperature, so does the approximated QFI. Above that temperature, both the mean field and the approximated QFI vanish. 

\begin{figure}[htp]
    \centering   
    \includegraphics[width=8.5cm]{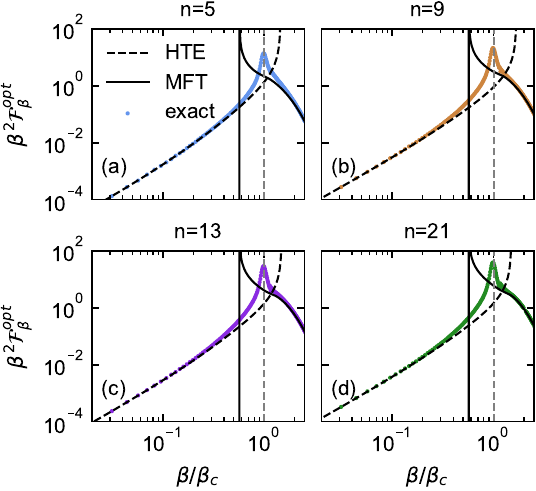}
    \caption{Optimal QFI of the phase probe as a function of inverse temperature relative to the critical point (vertical dashed line), extending the results in Fig.~\ref{local_pb_complt}(a). We separate the four cases corresponding to different cluster sizes $n$ into different diagrams (a)-(d), comparing the exact Monte-Carlo simulation data to the approximate predictions of the mean-field theory (solid lines) and the high-temperature expansion (dashed).
    \label{pb_lc_compare_plt}}
\end{figure}

In Fig.~\ref{pb_lc_compare_plt}(a)-(d), we compare the MFT-approximation for the QFI (black solid line) with the exact Monte-Carlo simulation results for four different cluster sizes. We find that \eqref{QFI_mft} predicts the temperature sensitivity well for subcritical temperatures, but fails at and above the phase transition. This is expected since above the critical point the magnetization vanishes. To study the QFI above the critical point, we use a high-temperature expansion \cite{stanley1987}.

\subparagraph{High-temperature expansion}
At high temperatures in the paramagnetic phase, thermal fluctuations are dominant and the lattice spins tend to be randomly oriented. The properties of the sample can be studied by performing an expansion of the relevant physical quantities in terms of the inverse temperature $\beta \ll 1/J$. Specifically, we must expand up to second order in $\tanh (J\beta)$ to arrive at a nontrivial result. The partition function expands as
\begin{eqnarray}
    \mathcal{Z} &=& \tr\left\{ e^{\beta \sum_{\langle ij \rangle} J \hat{\sigma}_z^i \otimes \hat{\sigma}_z^j } \right\} \\ \nonumber
    & =& \tr \left\{ \prod_{\langle ij \rangle}\Big[ \cosh{(\beta J)} + \sinh{(\beta J)} \hat{\sigma}_z^i \otimes \hat{\sigma}_z^j\Big]\right\} \\ \nonumber
    &\approx& \tr \left\{ \cosh^K(\beta J)\Big[ 1+ \tanh(\beta J) \sum_{\langle ij \rangle} \hat{\sigma}_z^i \otimes \hat{\sigma}_z^j \right. \\ \nonumber
    &&+ \left. \tanh^2(\beta J) \sum_{\langle ij \rangle < \langle kl \rangle} (\hat{\sigma}_z^i \otimes \hat{\sigma}_z^j) (\hat{\sigma}_z^k \otimes \hat{\sigma}_z^l)\Big]\right\} \\ \nonumber
    &=&  2^{N} \cosh^K(\beta J),
\end{eqnarray}
where the $K$ is the total number of pair bonds (i.e., interacting spin pairs) in the lattice. The double sum over pair bonds in the fourth line counts each pair only once.
To obtain the final form, notice that the first- and second-order terms in $\tanh (\beta J)$ in the third and fourth line vanish when taking the trace, since there are always at least two distinct spins contributing a factor $\tr \{\osigma_z^i \} = 0$. Here, the first non-vanishing contribution would come from the fourth-order summation over products of four pair bonds forming a closed loop between four adjacent spins~\cite{stanley1987}: $\la ij\ra, \la jk\ra, \la k\ell\ra, \la \ell i\ra$. 

\begin{figure}
    \centering
    \includegraphics[width=8.6cm]{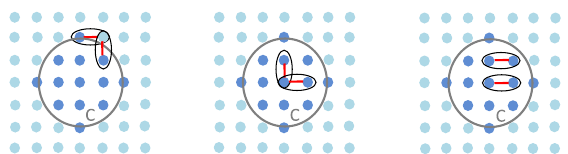}
    \caption{Construction of pair bond doublets in the Ising lattice. The cluster $C$ of spins interacting with probe (dark blue) are contained in a circle. Exemplary red lines symbolize pair bonds due to the Ising interaction between any two neighbouring lattice spins. From left to right, each marked doublet of pair bonds
    represents one possible contribution that is counted by the numbers $K_{22}$, $K_{23}$, $K_{24}$, respectively. In the depicted configuration there are $n=13$ cluster spins, $K_{12}=16$ intra-cluster bonds, and one obtains $(K_{22}, K_{23}, K_{24}) = (8,34,86)$. \label{construct_pair}}
\end{figure}

The phase probe's decoherence factor \eqref{charact_func} expands as
\begin{eqnarray}\label{eq:r_HTE}
    &&r(t, \beta) = \frac{1}{\cZ} \tr \left\{e^{\beta \sum_{\langle ij \rangle} J \hat{\sigma}_z^i \otimes \hat{\sigma}_z^j} e^{-igt \oZ_n }  \right\} \\
    &&\approx \frac{1}{\cZ}\cosh^K(\beta J) \Bigg[ \tr \left\{ e^{-igt \oZ_n } \right\} \nonumber \\
    && \quad + \tanh(\beta J) \sum_{\langle ij \rangle} \tr\left\{ (\hat{\sigma}_z^i \otimes \hat{\sigma}_z^j)e^{-igt \oZ_n }\right\}  \nonumber \\
    && \quad + \tanh^2(\beta J) \!\!\!\! \sum_{\langle ij \rangle < \langle kl \rangle} \!\!\!\! \tr \left\{ (\hat{\sigma}_z^i \otimes \hat{\sigma}_z^j)(\hat{\sigma}_z^k \otimes \hat{\sigma}_z^l) e^{-igt \oZ_n } \right\} \Bigg] \nonumber. 
\end{eqnarray}
Here, the first- and second-order terms in $\tanh (\beta J)$ no longer vanish due to the additional unitary generated by the cluster magnetization $\oZ_n$. Writing
\begin{equation}
    e^{-igt\oZ_n} = \cos^n (gt) \prod_{m=1}^n \left[ 1 -i\tan (gt) \osigma_z^m \right],
\end{equation}
we see that the first-order term in the third line of \eqref{eq:r_HTE} adds a term $\propto \tan^2(gt)$ for every pair bond $\la ij\ra$ whose both spins are contained in the cluster. Let $K_{12}$ denote the number of such intra-cluster pair bonds. 
For the second-order term in the fourth line of \eqref{eq:r_HTE}, non-vanishing contributions can come from doublets of pair bonds with at least two spins inside the cluster, as illustrated in Fig.~\ref{construct_pair}: $K_{22}$ counts the number of pair bond doublets with two spins inside and one mutual spin outside the cluster, $K_{23}$ counts doublets of intra-cluster pair bonds with one mutual spin, and $K_{24}=K_{12}(K_{12}-1)/2 -K_{23}$ counts doublets of intra-cluster pair bonds with no mutual spin. We arrive at
\begin{eqnarray}\label{eq:r_HTE2}
    &&r(t, \beta) = \cos^n(gt) \bigg\{ 1 -\tanh(\beta J)\tan^2(gt) \nonumber \\
    &&\quad \times \left[ K_{12} - \tanh(\beta J) \left(K_{22} + K_{23} - \tan^2(gt) K_{24}  \right)\right] \bigg\}. \nonumber 
\end{eqnarray}

The QFI of the phase probe with respect to the inverse temperature can once again be given analytically. For a compact expression in the high-temperature regime, we now approximate $\tanh(J \beta) \approx J \beta$ and arrive at
\begin{eqnarray}
    \cF_{\beta}^{(\mathrm{HTE})} &=&  \frac{J^2 \Lambda^2(t) \tan^4 (gt) \cos^{2n}(gt)}{1 - \left[ J \beta \Lambda(t) \tan^2(gt) - 1\right]^2 \cos^{2n}(gt)}, \label{eq:QFI_HTE} \\
    \Lambda(t) &=& K_{12} +2J \beta \left[K_{22} +K_{23} -K_{24} \tan^2 (gt) \right]. \nonumber 
\end{eqnarray}
The phase probe thus picks up temperature information through the correlations between at most four lattice spins. 
In Fig.~\ref{pb_lc_compare_plt}, the dashed lines depict the HTE \eqref{eq:QFI_HTE} of the QFI, which matches the exact results for $\beta \ll \beta_c$ in the paramagnetic phase. 

We conclude that the actual critical behaviour and the peak value of the phase probe's temperature sensitivity cannot be accurately described by either analytical approximation to the QFI. The MFT fails due to the buildup of strong long-range thermal fluctuations, while the HTE fails due to the buildup of long-range multi-spin correlations in the critical Ising lattice. Moreover, the finite peak value of the QFI cannot be captured by continuous models that describe the phase transition in the thermodynamic limit, necessitating numerical simulations. 

\section{Conclusion} \label{conclusion}

In this work, we considered non-invasive quantum thermometry of finite, strongly interacting Ising spin systems at temperatures around their critical phase transition point. We demonstrated that the non-Markovian dephasing of a quantum probe in the vicinity of such a critical system picks up thermal spin fluctuations and thus reaches peak temperature sensitivity at the critical point. Consequently, one can estimate the system temperature by monitoring the probe's phase coherence over time, and one achieves the highest estimation precision (i.e., minimum relative error, as measured by the QFI) at the critical temperature---applying critical quantum metrology \cite{Rams2018, Garbe2020, Chu2021, Raffaele2023} for temperature estimation.

To exemplify the critical sensitivity enhancement in an instructive analytical manner, we first performed a case study of phase thermometry on the Curie-Weiss model of a fully connected spin network in the thermodynamic limit. The QFI of a single spin probe exhibits a singularity at the critical temperature, and the probe dynamics changes from an oscillatory to a non-oscillatory decay of phase coherence as the system transitions from the ferromagnetic to the paramagnetic phase. The achievable sensitivity clearly surpasses that of a local single-spin measurement of the lattice.

For our main results, we performed Monte-Carlo simulations on a $20\times 20$ Ising lattice with ferromagnetic nearest-neighbour coupling, showing a finite critical sensitivity peak of the probe that depends on the number of lattice spins it interacts with. In the critical range of temperatures, analytic approximations such as mean-field theory for the ferromagnetic phase and the high-temperature expansion for the paramagnetic phase do not accurately capture the thermal spin fluctuations leading to the strongest response of the probe. Accordingly, phase thermometry constitutes a suitable, non-invasive metrology scheme to access the theoretically elusive critical regime of strong fluctuations in interacting many-body systems.
Future research could explore phase probes for temperature estimation in finite samples of Heisenberg-type spin systems, for example. Bayesian phase thermometry schemes with optimal, ancilla-augmented probe states could also be considered \cite{Bavaresco2023}. Moreover, the spreading of Fisher information from a target location across the spin lattice could be considered for remote sensing and parameter estimation tasks.

\acknowledgments
The University of Siegen is kindly acknowledged for enabling our computation through the \texttt{OMNI} cluster. 
This work was supported by 
the Deutsche Forschungsgemeinschaft (DFG, German Research Foundation, project numbers 447948357 and 440958198), 
the Sino-German Center for Research 
Promotion (Project M-0294), 
the ERC (Consolidator Grant 683107/TempoQ), 
and the German Ministry of Education 
and Research (Project QuKuK, BMBF 
Grant No. 16KIS1618K).

%

\end{document}